# The Low Energy X-ray telescope (LE) onboard the Insight-HXMT astronomy satellite

Yong Chen[1*], WeiWei Cui[1], Wei Li[1], Juan Wang[1], YuPeng Xu[1], FangJun Lu[1], YuSa Wang[1], TianXiang Chen[1], DaWei Han[1], Wei Hu[1], Yi Zhang[1], Jia Huo[1], YanJi Yang[1], MaoShun Li[1], Bo Lu[1], ZiLiang Zhang[1], TiPei Li[1,2,3], ShuangNan Zhang[1,2], ShaoLin Xiong[1], Shu Zhang[1], RongFeng Xue[1,4], XiaoFan Zhao[1], Yue Zhu[1], YuXuan Zhu[1,4,] HongWei Liu[1], YiRong Yang[1], Fan Zhang[1]

[1]Key Laboratory for Particle Astrophysics, Institute of High Energy Physics, Beijing 10049, China
[2] University of Chinese Academy of Sciences, Chinese Academy of Sciences, Beijing 100049, China
[3] Tsinghua University, Beijing 100084, China;
[4]College of Physics, Jilin University; Changchun 130023, China

**Abstract:** The low energy (LE) X-ray telescope is one of the three main instruments of the Insight-Hard X-ray Modulation Telescope (*Insight*-HXMT). It is equipped with Swept Charge Device (SCD) sensor arrays with a total geometrical area of 384 $cm^2$ and an energy band from 0.7 keV to 13 keV. In order to evaluate the particle induced X-ray background and the cosmic X-ray background simultaneously, LE adopts collimators to define four types of Field Of Views (FOVs), i.e., $1.6° \times 6°$, $4° \times 6°$, $50~60° \times 2~6°$ and the blocked ones which block the X-ray by an aluminum cover. LE is constituted of three detector boxes (LEDs) and an electric control box (LEB) and achieves a good energy resolution of 140 eV@5.9 keV, an excellent time resolution of 0.98 ms, as well as an extremely low pileup (<1%@18000 cts/s). Detailed performance tests and calibration on the ground have been performed, including energy-channel relation, energy response, detection efficiency and time response.

# 1 Introduction

The Low Energy (LE) X-ray telescope [1-2] is one of the science instruments onboard the Hard X-ray Modulation Telescope (HXMT) astronomy satellite, dubbed as *Insight*-HXMT after launch on June 15, 2017. The scientific objectives of LE focus on the scanning and pointed observations of the X-ray sources in the soft X-ray band (0.7-13keV).

Unlike Chandra and XMM-Newton X-ray observatories that utilize grazing incidence telescope to focus photons [3], LE uses collimators to shield the photons outside its field of view (FOV).

The Swept Charge Devices (SCDs) of LE can work in a continuous readout mode [4], recording the energy information and the time of arrival (TOA) of an incident X-ray photon to achieve a higher time resolution, which are different from the conventional Coupled Charged Devices (CCDs) [5]. In total, 24 SCD modules (96 channels) are assembled in LE with a total geometrical area of 384 cm$^2$, which is the largest among the X-ray CCD sensor arrays in space missions. Together with the good energy and time resolutions, LE is expected to play an important role in soft X-ray astronomy. The summary of the LE characteristics is shown in Table 1.

Onboard *Insight*-HXMT, LE has been operating on-orbit for more than one year and has observed tens of X-ray sources. Some essential scientific results have already been obtained gradually (e.g. [6-7]).

**Table 1** Summary of the LE characteristics

| Item | Performance |
|------|-------------|
| Detector area | 384 cm$^2$ |
| Detector type | Swept Charge Device (SCD) |
| Energy range | 0.7-13 keV |
| Collimator FOV | 1.6° × 6°, 4° × 6°, blocked FOV, 50~60° × 2~6° (trapezoidal) |
| Energy resolution | 140 eV@5.9 keV |
| Time resolution | 0.98 ms |
| Sensor operating temperature | -80℃~ -30℃ |

# 2 Instrument Description

## 2.1 Overview

LE contains three identical detector boxes (LEDs) and one electric control box (LEB). As shown in Figure 1, these three LEDs are aligned and mounted upon the payload's main plate. Therefore, during the scanning observation of HXMT, the signals from the X-ray sources will be modulated by the collimator, and the image can be obtained by a reconstruction technique, e.g. the Direct Demodulation method [8-9].

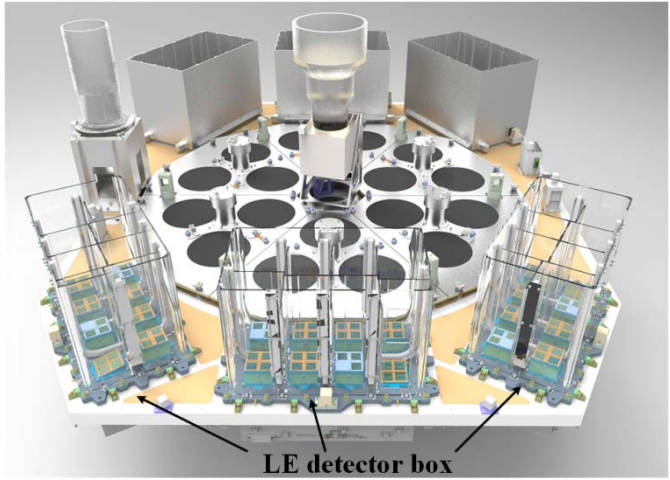

**Figure 1** Structure of LE layout on HXMT

Figure 2 shows a block diagram of the LE electronics. The SCD arrays are located in the LEDs, which convert proportionally the energy of X-ray photons to analog voltage signals. The electronics in the LED processes the SCD output signal and converts it into digital representation to generate the event data containing both the energy information and the TOA of an incident X-ray photon. The event data are sent to the LEB through the low-voltage differential signaling (LVDS) interface.

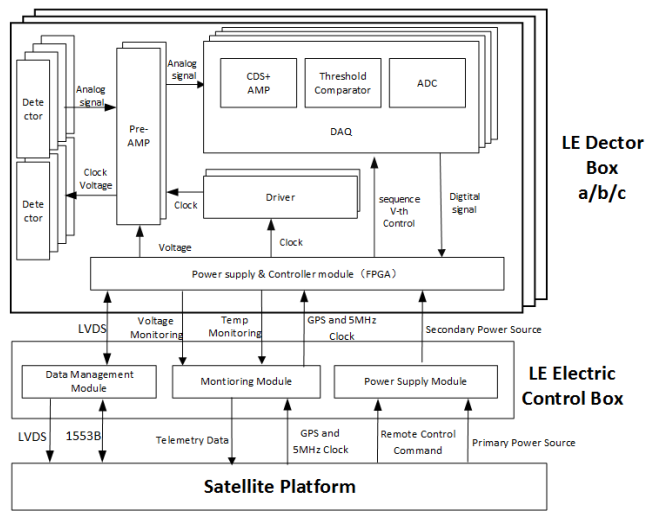

**Figure 2** Block diagram of the LE electronics

The functions of LEB are data management, communication as well as power supply and distribution. Moreover, it is capable of automatic adjustment of the on-board noise threshold and on-line count rate and spectrum processing. It is also the interface between the LEDs and the satellite platform, through which telemetry, remote control as well as the instruction and data transmission can be carried out.

Due to the fact that only below -30℃ can the SCD sensors have optimal performances [10, 13], (see also section 4.2 of this paper for detail), the LEDs are directly exposed to space, so that the heat can be effectively radiated. In addition, radiative cooling boards are firmly connected to the detectors to increase the cooling efficiency, and the detectors are insulated from the electronics that have relatively high power consumption [10].

## 2.2 LED

As shown in Figure 3, a LED contains two parts, i.e. the upper part and the lower part, which are mounted above and below the payload's main plate (as shown in Figure 1) respectively. Flexible cables are adopted for electrical connection between these two parts. Each LED consists of eight SCD modules, two types of collimators, visible light blocking filters, anti-contamination films, heat pipes as well as several thermal and mechanical supporters. The lower part of the LED mainly contains the electronics.

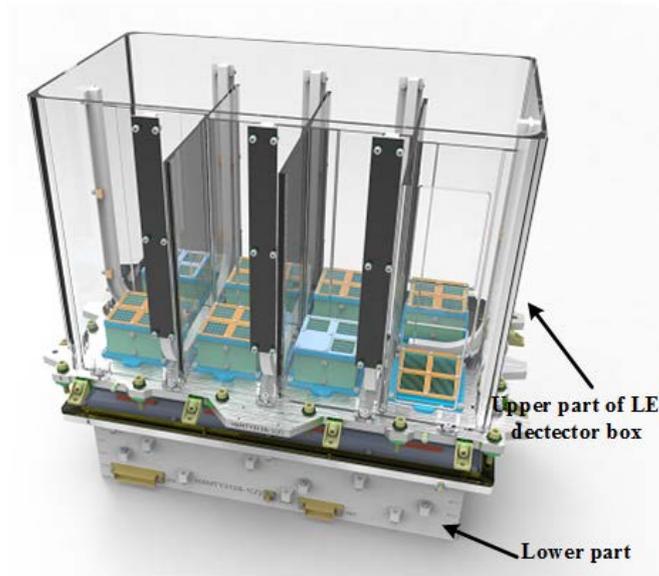

**Figure 3** The structure of LED

### 2.2.1 SCD Sensor

As a special type of CCD, SCD [11] features processing large geometrical area through a configuration electrically similar to a linear CCD. Whilst the characteristic is desired for LE application, the area for a monolithic chip should be enlarged and the readout speed accelerated to meet the power budget requirements with an acceptable time resolution.

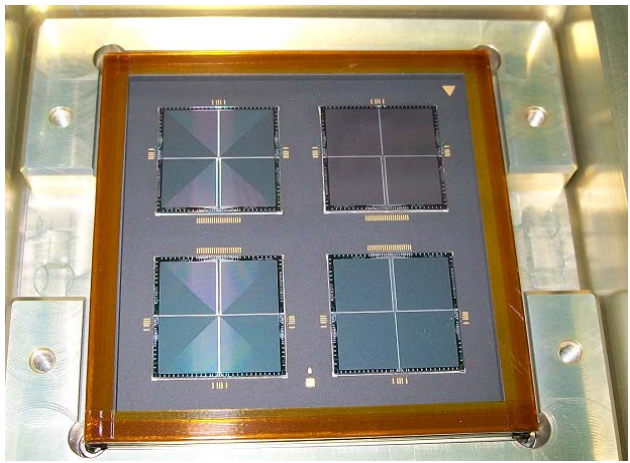

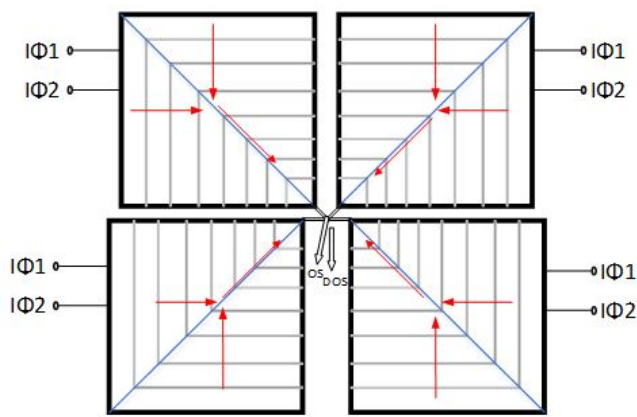

**Figure 4** Upper panel: Photo of an LE detector module. It contains 4 CCD236 chips with a total geometrical area of 16 cm². Lower panel: Schematic of CCD236. The red arrows indicate the charge transfer directions.

CCD236 is the second-generation SCD [12], which has been developed by e2v Inc. from UK under full considerations of the requirement of LE. It has a total effective area of 4 cm², comprising four quadrants. In each quadrant, the L-shaped electrodes guide the charge towards the diagonal first and then to a common readout amplifier in the central region. Therefore, the signals from different pixels could be transferred into the readout amplifier in the same time and the location information is almost lost, which is different from that of a conventional CCD. A dedicated dummy amplifier, which has the same readout circuit structures but is not connected to the effective pixels, is provided to suppress the common mode interference by means of subtracting its amplitude from the readout amplifier. In the continuous readout mode, the total readout time is only 0.98 ms with a guaranteed energy resolution. The threshold energy is well below 1 keV.

The implementation of the sensor for LE involves the packaging of four CCD236 devices into a single module, as shown in Figure 4. As a consequence, the effective area for a single module is 16 cm².

A series of experiments and tests, including the usual environment tests and particle irradiation experiment etc, were performed to ensure that LE will work properly in the complex space environment. The CCD236 operates smoothly in all these tests [13-14].

### 2.2.2 Collimators

Several FOVs are adopted by LE to fulfill various scientific observation goals. To accomplish that, a collimator is mounted on the top of every SCD sensor module to limit its FOV.

Multiple FOVs are designed for LE to measure the cosmic X-ray background and the particle induced background. The particle background should be removed for all types of the observations. The cosmic X-ray background should be eliminated in the observations of point sources, but its properties are also the scientific objectives of *Insight*-HXMT. As the particle background is not sensitive to the FOV and the influence of the diffuse X-ray background is proportional to FOV, we used the blocked FOV to mainly measure the particle background, narrow FOV ($1.6° \times 6°$, FWHM) mainly for point source observation, and the wide FOV ($4° \times 6°$, FWHM) to mainly determine the diffuse X-ray background. A similar design was also used by the X-ray satellite HEAO1-A2 to distinguish the two kinds of background [15]. The survey collimators have trapezoidal FOVs and are used to survey the large sky area in the scanning mode. Moreover, there are two kinds of sub-FOVs (FOV1 & FOV2) which are slightly different, because of the different heights of the collimator walls over different SCD chips (Figure 6 lower panel). The smallest size of the top-line of the trapezoid is 2°(FOV1) and the largest size of baseline is 6°(FOV2). The heights of trapezoids of the two FOVs are 50°(FOV1) and 60°(FOV2), respectively.

Figure 5 shows the overhead view of the collimators assembled in an LED. Among the 32 sensor chips inside, there are 20, six and two sensor chips possessing narrow, wide and blocked FOVs, respectively, and the remaining four sensor chips are mounted beneath the survey collimators. All collimators are made of aluminum. The thickness of the collimator wall is about 120 μm and can entirely block the X-ray with the energy less than 10 keV outside the FOVs. The off-axis response can be simply evaluated as F =（Δ α/α）( Δ β/β), where α and β are the two dimension angles of the FOVs, Δ α and Δ β are the off-axis angles of the two dimensions, F is the value of the off-axis response.

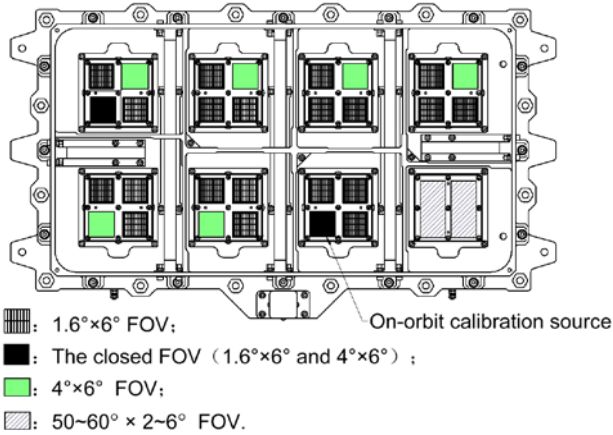

On-orbit calibration source

▦ : 1.6°×6° FOV；
■ : The closed FOV（1.6°×6° and 4°×6°）；
🟩 : 4°×6° FOV；
▨ : 50~60° × 2~6° FOV.

**Figure 5**. Layout of collimator FOVs in the LED

A wide FOV and a narrow FOV among the seven normal collimators are blocked, which is implemented by aluminum cover with a thickness of 1 mm. Furthermore, for on-orbit calibration purpose, a Fe[55] radioisotope (0.1~1μCi) is mounted in the blocked FOV.

Figure 6 shows the configuration of both the normal and survey collimators. The optical blocking filters (OBFs) mechanically strengthened by Ni frames are immobilized on the collimators. With the protection of special-shaped ventilation holes distributed on the top of the collimators, the optical blocking filters can survive the severe airflow impact during the satellite launch. The wide and the narrow FOVs, with dimensions of 3.7mm×5.47mm and 1.47mm×5.47mm, respectively, are machined by the wire-electrode cutting technique. The normal collimators are attached around with 0.2 mm thick tantalum plates to reduce the background induced by gamma-rays and the charged particles in space. Besides, the anti-contamination films (ACFs) and bottom caps are assembled together at the bottom of the collimators. Furthermore, the survey collimator has a shortened height and a wedged top surface to achieve a large FOV.

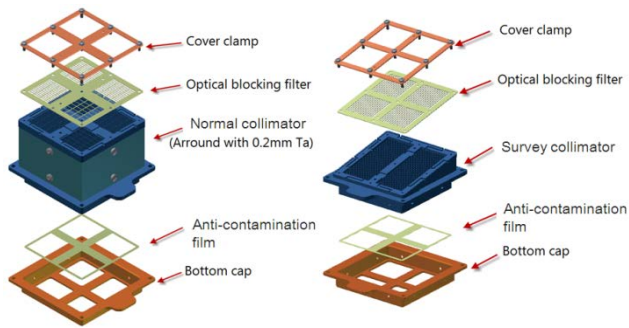

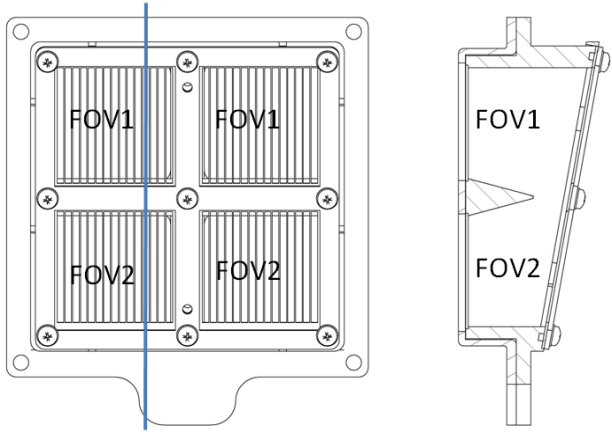

**Figure 6.** Upper panel: Exploded view of LE collimator (left: normal collimator, right: survey collimator). Lower panel: sub-FOVs (FOV1 & FOV2) of the survey collimator. A section view along the blue line of the survey collimator (left) is showed in the right.

The surface of LE collimator is processed with black anodic oxidation treatment. Figure7 shows the photograph of the blackened collimators.

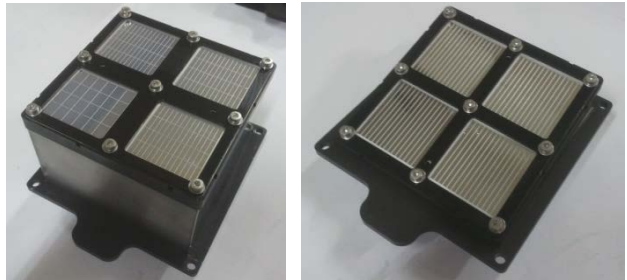

**Figure 7.** Photograph of LE collimators (left: normal collimator, right: survey collimator)

The aperture opening ratio of the collimator (the proportion of hole area) which is influenced by processing precision, is directly related to the effective geometrical area. Thanks for the low-speed wire cutting technique. The processing precision of the collimator is less than 0.01 mm. The aperture opening ratios of LE collimators are summarized in Table 2.

**Table 2** LE collimator aperture opening ratios

| FOV | Narrow | Wide |
|---|---|---|
| Collimator hole number | $14 \times 4$ | $6 \times 4$ |
| Aperture opening ratio in practice | 91.6% | 95.9% |

### 2.2.3 OBF and ACF

Since SCD sensor is highly sensitive to visible and ultraviolet light, it is necessary to filter out these undesired radiations (lights). The OBF is mounted onto the collimator to block ultraviolet and visible lights while remaining transparent to soft X-rays. Also, it meets the requirements such as mechanical strength, acoustic vibration and atomic oxygen erosion [16-17].

The OBF has a sandwiched structure composed of aluminum foil, polyimide film and nickel frame. The aluminum foil with a thickness of 100±10 nm is deposited on both sides of the polyimide film. The polyimide film with a thickness of 400±20 nm could support the aluminum foil and thus greatly improve its mechanical properties. The nickel frame with a thickness of about 80μm is applied to support the OBF and provide the mounting interface.

Considering that LE uses both normal and survey collimators, there are also two types of OBFs. Most of their parameters are the same except for the grid pattern and the line width of the nickel frame (Figure 8).

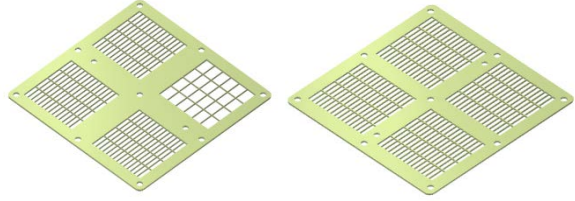

**Figure 8** Diagram of **OBFs (left: for normal collimator, right: for survey collimator)**

The UV-visible transmittance experiment was carried out with a UV-VIS spectrometer. From the results shown in Figure 9, the OBF containing a single aluminum layer of 60 nm (the black curve) shows a UV-visible transmittance lower than 4.5%. Whereas those with aluminum foil thicker than 60 nm (120 nm and 180 nm, represented with blue open rectangle and red open triangle respectively in Figure 9) have transmittance almost overlapped with the baseline. According to those results, the aluminum foil thickness used for the OBF is selected to be 200 nm. It is worth noting that the measuring error of UV-VIS spectrometer is about $1 \times 10^{-5}$ and it can hardly measure the exact transmittance of OBF with an aluminum foil thicker than 120 nm in practice. The calculated transmittance of an OBF (Al/PI/Al = 100/400/100 nm) in UV-VIS band is around $10^{-8}$ (Figure 10).

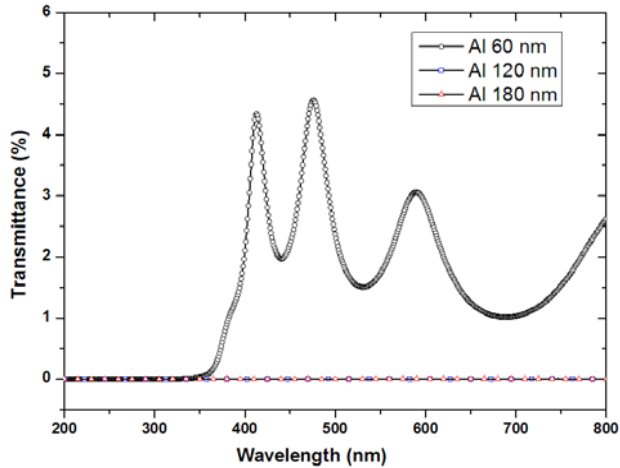

**Figure 9** The transmittance of optical blocking filters in UV-VIS band

The X-ray transmittance of the OBF is crucial to assess its performance. Figure 11 shows the transmittance of the OBF (Al/PI/Al = 100/400/100 nm) in soft X-ray band. The experimental curve is close to that of the theoretical fitting. The measured soft X-ray transmittance is as high as about 78% at 1 keV.

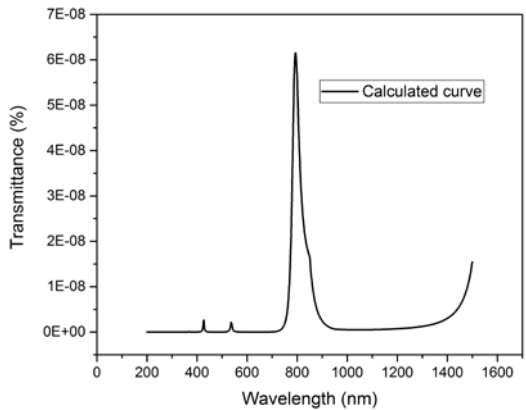

**Figure 10** The calculated transmittance of optical blocking filters in UV-VIS band (Al/PI/Al = 100/400/100 nm).

Space irradiation is one of the main factors that affects the performance of the OBF. We therefore performed proton (10 MeV) irradiation experiments for three OBFs (Al/PI/Al=100/400/100 nm) with a dose of $10^{10}$ protons/cm$^2$, the expected irradiation dose of LE in its 4-year operation. Figure 11 shows the soft X-ray transmittance of the OBF with and without irradiation for a total of 6 OBFs. We should note that due to limited chance of the synchrotron experiment for the X-ray transmittance measurement, we divided the six OBFs into two sets, each has three OBFs. Only with one set (irradiated set), we carried out the proton irradiation experiments. The residuals of averaged transmittance of original and irradiated OBF deviated from the simulated curves are shown in the lower panel of Figure 11. There are only slight differences in transmittance between the residuals of the two sets, however, they are well within the individual scatters of the transmittance for the whole sample. It suggests that proton irradiation has negligible influence on the transmittance of the OBF. The residuals also showed a declining trend (Figure 11 lower panel), which may indicate that there are some defects, e.g. pinholes, wrinkles, non-homogenous in thickness and high-Z element contaminations, in the OBFs. For example, the high-Z element contamination can reduce the transmittance at high energy efficiently, while the pinholes can increase the transmittance significantly at low energy but negligibly at high energy.

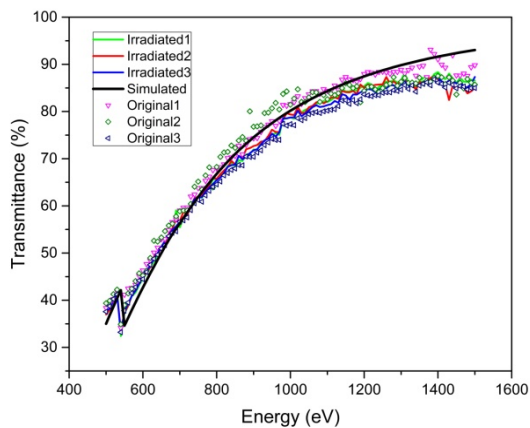

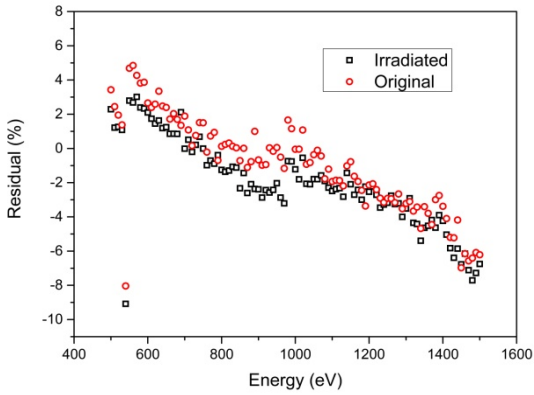

**Figure 11** Upper panel: the transmittance in soft X-ray band of 6 LE OBFs with three of them irradiated by proton. Bold black curve corresponds to the simulated transmittance. The residuals of averaged transmittance of original and irradiated OBF deviated from the simulated curves are shown in the lower panel.

Contaminated metal particulates could cause short circuit problem to the SCD sensor, and so an ACF is mounted beneath the collimator to avoid this problem. Different from the OBF, the ACF is composed of a layer of 300 nm polyimide film, which is supported by nickel frame without grids and AL layer to minimize the soft X-ray blocking. As shown in Figure 12, a press frame flange and a collimator flange are set up to mount the ACF.

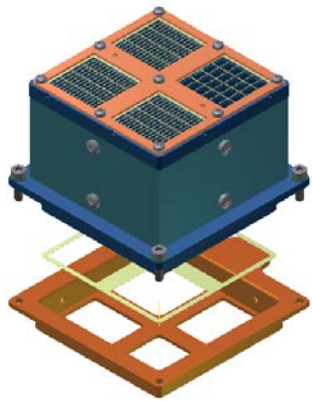

**Figure 12** Exploded view of the collimator and the ACF

### 2.2.4 Electronics of LED

The electronics of each LED functions for data acquisition. It adopts modular design methodology, including the analog front-end module, data acquisition module, driver module and power supply and control module.

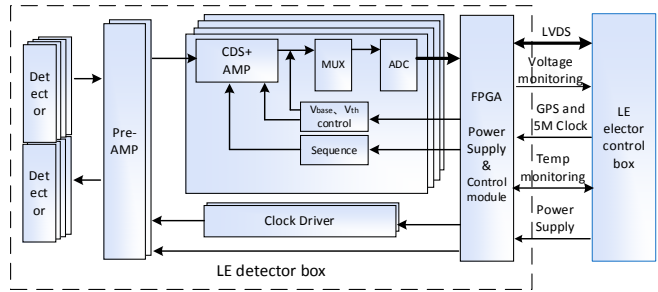

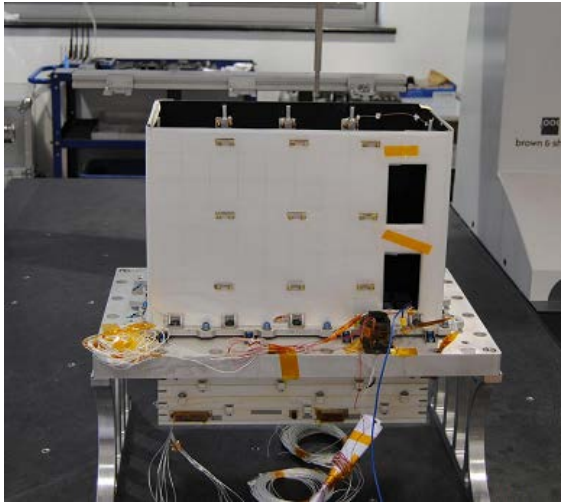

**Figure 13** The block diagram of the electronics (upper) and photograph (lower) of LED.

The details of the electronics are shown in Figure 13, there are two identical submodules for both the preamplifier circuit and the driver module, whereas four identical submodules for data acquisition module. In addition, the control module is also composed of two identical submodules, but each is a cold backup.

The DAQ module converts the detector voltage signal into a signed 12-bit digital signal with a maximum sample rate of 100 k sample/s. An Actel anti-fuse FPGA is adopted in the DAQ module to achieve a good performance on the suppression of the single event effects in the space.

## 2.3 LEB

Each LEB is made up of three parts, including data management module (DMM), sensor monitoring module (SMM) and power supply module (PSM), as shown in Figure 14. DMM and PSM have both primary and standby sets.

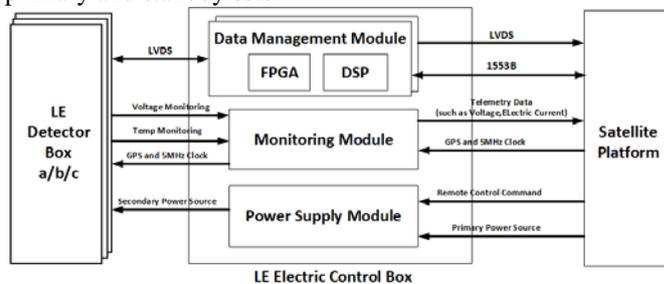

**Figure 14** Block diagram of each LEB

DMM receives both science data and engineering data generated by the three LEDs. After the insertion of auxiliary identifiers, these data are packaged into a packet in CCSDS format, and then transmitted to the data transmission system of the platform through LVDS interface. The science data are also processed onboard into specific data of count rate and energy spectrum, which are finally sent to the data management system of the platform through the 1553B interface. Besides, various bus instructions are received by the 1553B interface, and sent to the corresponding execution units after parsing.

DMM monitors such status parameters as current, voltage, temperature, etc., which are transmitted to the earth through telemetry. To enhance the detection efficiency, all the noise thresholds for the 96 CCD channels can be automatically adjusted by each LEB with the temperature variation feedbacks from each CCD module. The period for automatic threshold-adjustment is 64 s, and the threshold accuracy is about 3.3 eV. There are 16 temperature monitoring points with effective range from -100 ℃ to 40 ℃. It is worth mentioning that, thresholds can be optimized according to the CCD performance.

PSM converts the primary 28 V supply voltage from the platform into a series of secondary voltages through DC-DC converters, and outputs four DC voltages (+30 V/+7 V/±6 V) required by all LEDs.

In addition, LEB plays critical roles in receiving and forwarding such data to the corresponding executive units as remote control instructions, GPS Pulse per second, UTC time information and 5 MHz clock signals. The data management function can be optimized onboard. A photo of the flight module of LEB is shown in Figure 15. The maximum data acquisition and processing rate of photon events is 44,000 counts/s.

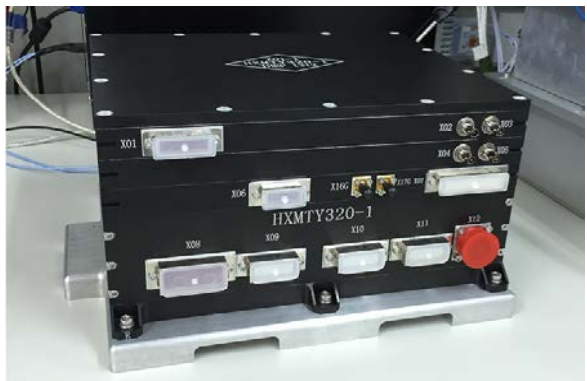

**Figure 15** Photo of a LEB

# 3 Operation Mode

LE has various operation modes. For X-rays with energy above the pre-set thresholds, their arriving time, energy and channel numbers are recorded by the electronics. The thresholds can be remote-controlled. This kind of trigger is also described as "over threshold trigger (OTT)". Besides this, the other trigger mode for LE is the "forced trigger (FT)", which records the amplitude (energy) of each channel every 32 ms. The goal of FT is to obtain the noise peak profile for each detector.

There are two kinds of data acquisition modes of LE, the normal mode and the Southern Atlantic Anomaly (SAA) mode. Under the normal mode, the events from OTT and FT are transferred to the LEB. But only the FT events are transferred to the LEB when LE switches to SAA mode.

There are also two kinds of readout modes for CCD236, the continuous readout mode and the long exposure readout mode. When CCD236 is under continuous readout mode, the charge in the pixel is transferred one by one without break, and recorded continuously. In the long exposure mode, the charge in a pixel is suspended for a certain duration from several ms to several hundred ms to collect more dark charge, and then the whole charge in one chip will be continuously transferred and recorded. The long exposure mode is used to check the dark charge pattern and to measure the time response of a CCD236.

Since there are two kinds of pixel rate, 100 kHz and 50 kHz, for the sensor, 16 operation modes are designed as shown in Table 3.

**Table 3** Operation mode of LE

|  | Readout mode | Pixel rate | note |
|---|---|---|---|
| 1 | continuous readout | 100 kHz | Main mode |
| 2 | continuous readout | 50 kHz | Secondary mode |

| 3 | long exposure readout 2 ms | 100 kHz | |
|---|---|---|---|
| 4 | long exposure readout 4 ms | 50 kHz | |
| 5 | long exposure readout 5 ms | 100 kHz | |
| 6 | long exposure readout 10 ms | 50 kHz | |
| 7 | long exposure readout 10 ms | 100 kHz | |
| 8 | long exposure readout 20 ms | 50 kHz | |
| 9 | long exposure readout 20 ms | 100 kHz | Less commonly used |
| 10 | long exposure readout 40 ms | 50 kHz | |
| 11 | long exposure readout 50 ms | 100 kHz | |
| 12 | long exposure readout 100 ms | 50 kHz | |
| 13 | long exposure readout 100 ms | 100 kHz | |
| 14 | long exposure readout 200 ms | 50 kHz | |
| 15 | long exposure readout 200 ms | 100 kHz | |
| 16 | long exposure readout 400 ms | 50 kHz | |

When in orbit, LE will mostly work under continuous readout mode with 100 kHz pixel rate.

# 4 Calibration

## 4.1 Energy-Channel relationship

The energy-channel (E-C) relationship can be evaluated by integral nonlinearity, which can be described as below:

$$INL = \frac{\left| E_i - \hat{E}_i \right|_{\max}}{E_{\max} - E_{\min}} \times 100\% \quad (1)$$

where $E_i$ and $\hat{E}_i$ are the central positions of the arriving X-ray peaks through Gauss fitting and the positions obtained from E-C curves, respectively. $\left| E_i - \hat{E}_i \right|_{\max}$ is the maximal deviation among the above two kinds of positions.

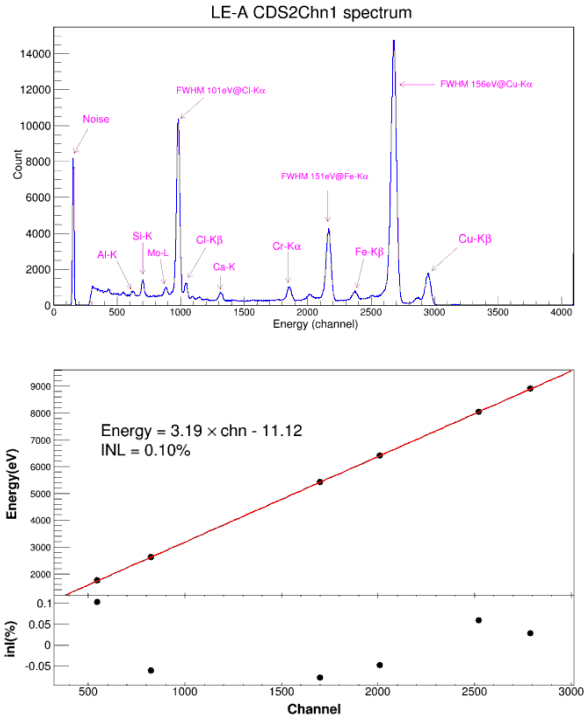

**Figure 16** An example spectrum (upper) and the integral linearity (INL) (lower) obtained in the thermal test of LE flight model

Figure 16 shows an example spectrum obtained in the thermal test of LE flight model, where a copper target is bombarded by photons from an X-ray tube. The differences of the amplitudes of each readout channel are within ±1%. The temperature-dependent Cu-Kα (8.048 keV) signal amplitude curves of 32 readout channels are shown in Figure 17.

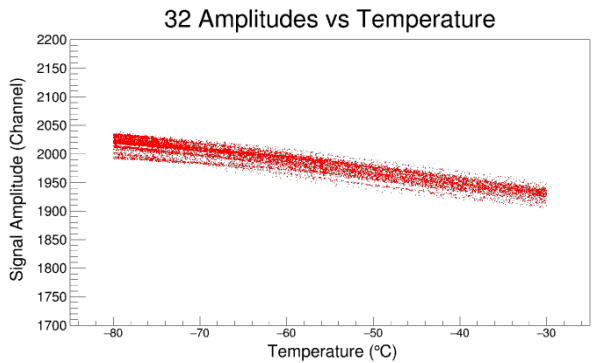

**Figure 17** Temperature-dependent Cu-Kα signal amplitude curves of 32 channels, obtained from the LE flight model

### 4.2 Readout noise and energy resolution

Once in the satellite orbit, the working temperatures of LE detectors are determined by the Sun angle, the infrared radiation of the Earth and the readout modes of LE. Both simulations and experiments show that the working temperatures of the sensors vary within the range of -80℃~-30℃. Consequently, the positions of noise peaks will shift, so will the readout noise and energy resolution during observations.

For one of the three LEDs, the merged spectrum of 32 channels is similar to the spectrum of each single channel. As shown in Figure 18, the full-width-half-maximum (FWHM) of the merged spectrum is 158 eV@8.0 keV, degrading slightly compared with that from a single channel.

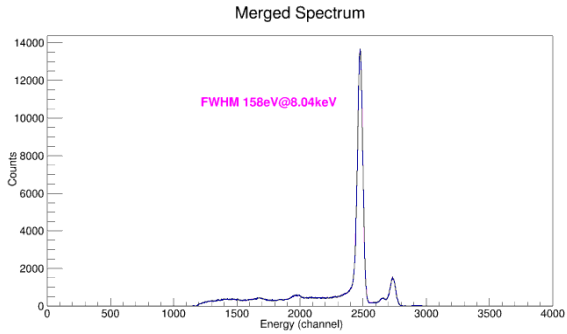

**Figure 18** The merged spectrum of 32 channels, obtained from the flight model

The temperature effect on the readout noise and FWHM of LE are shown in Figure 19 and Figure 20 respectively, where the temperature of sensor changes from -80℃ to -30℃. When the temperature is below -50℃, the readout noise is almost a constant, and the differences between channels are negligible. Above -50℃, the readout noise increases quickly with temperature, due to the increasing dark current at relatively high temperatures. The differences among channels become also significant at high temperatures. The relation between energy resolution and temperature is similar to that of the readout noise, the energy resolution is almost constant below -40℃. These results indicate that some of the SCDs may have some kind of defects, e.g. the cosmetic defect, and the noise as well as the energy resolution would be affected by the defects severely at high temperature.

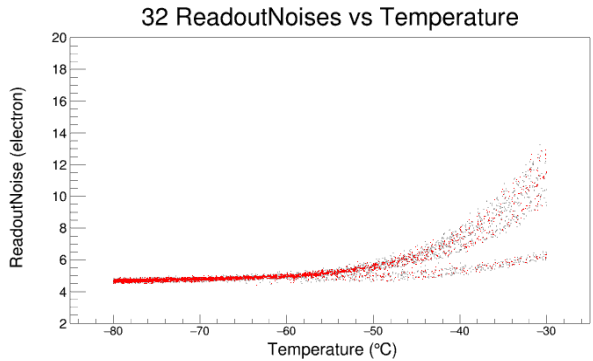

**Figure 19** Temperature-dependent readout noise curves of 32 channels, obtained from the flight model

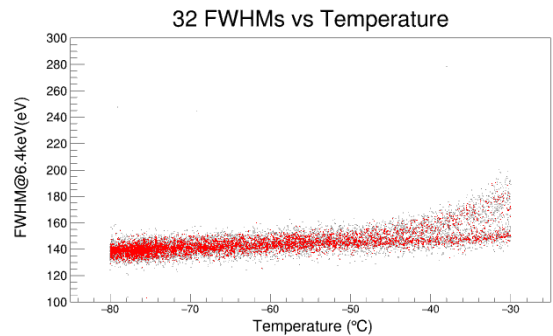

**Figure 20** Temperature-dependent energy resolutions (FWHM) of the 32 detector channels at the Fe Kα energy obtained from the flight model

### 4.3 Energy response

The energy responses have been measured with double-crystals monochromator that is described in details in the paper about the Medium Energy X-ray telescope of *Insight*-HXMT [18-19]. Figure 21 shows a typical energy response to 5.5 keV monochromatic X-ray, which mainly includes the noise peak, the raw spectrum, and the single event spectrum.

The split events are generated when an X-ray photon interacts with two or more adjacent pixels. Since the charge transfer and readout are continuous, the two or more events are recorded with continuous arriving time in data, which can be used to select the split events. The split events are less than 20% of all raw events. Also, split events can be restored by adding the adjacent events together, to increase the effective area. The split event ratio is almost independent of temperature and shows a slight increasing trend with the increase of the energy of photons, which is due to that the high energy photons usually create electrons in much deep region inside the detector.

The differences between the raw spectrum and the single event spectrum are showed in Figure 21. From the energy responses to X-rays with various energies, the redistribution matrix file (RMF) can be generated by an interpolating technique, as illustrated in Figure 22.

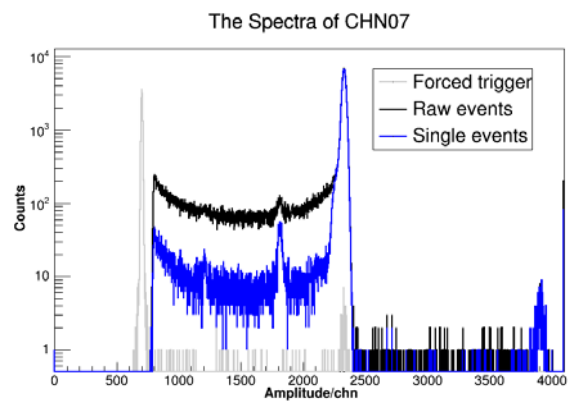

**Figure 21** LE's response function at -50.3℃ with 5.5 keV monochromatic X-ray.

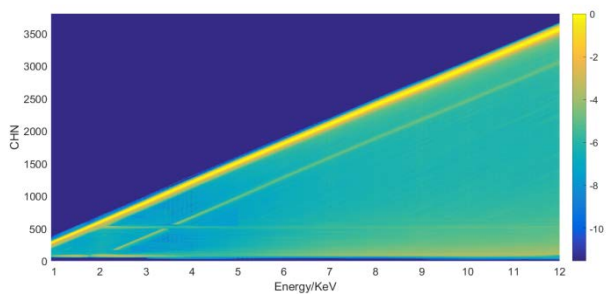

**Figure 22** An example RMF of LE.

### 4.4 Quantum efficiency

The relative efficiency of LE has been evaluated with the electron-beam ion trap (EBIT) facility in Shanghai [20]. The comparison of the measured energy spectra between LE sensor operating at -100℃ and the standard sensor (High Purity Germanium (HPGe) sensor) is shown in Figure 23. Both spectra contain six peaks whose energies are 6.7, 8.3, 9.1, 10.4, 11.9, and 13.3 keV, respectively. In the LE sensor spectrum, the counts of 11.9 and 13.3 keV peaks are less than those of other peaks, which is due to the fact that the efficiency of CCD236 decreases significantly at high energies. Comparing the total counts of these peaks between the CCD 236 and HPGe spectra, the relative quantum efficiency of CCD236 to HPGe, as shown in Figure 24, can be obtained. For energy below 3.5 keV, the quantum efficiency of CCD236 is much higher, because the entrance window of the HPGe sensor is much thicker.

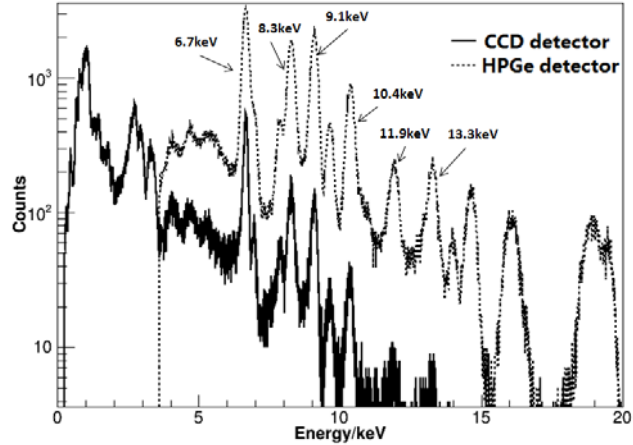

**Figure 23** The energy spectra of the radiation from an EBIT by CCD236 (-100℃) and HPGe [20].

The theoretical efficiency curve of CCD236 can be calculated using its depletion layer thickness of 50 µm. According to the simulated and normalized results, the absolute efficiency at 6.7 keV is 0.698, and the efficiencies at other energy points could be obtained through the same methods. Comparison between experimental efficiency and the theoretical efficiency is presented in Figure 24.

It is difficult to calibrate the SCD efficiency at the low energy range directly. One can obtain the efficiency by extrapolating the above data to the low energy range [20]. In addition, it may be more convenient to calibrate the LE effective area on orbit as done in [21].

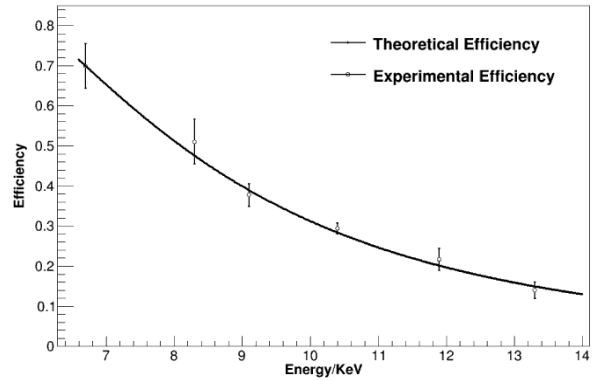

**Figure 24** Comparison of the experimental efficiency with the theoretical relative efficiency of CCD236.

The calculated total on-axis effective area of LE (including both the narrow and wide FOV SCDs) is showed in Figure 25, in which the detector efficiency and the filter transmittance are already considered.

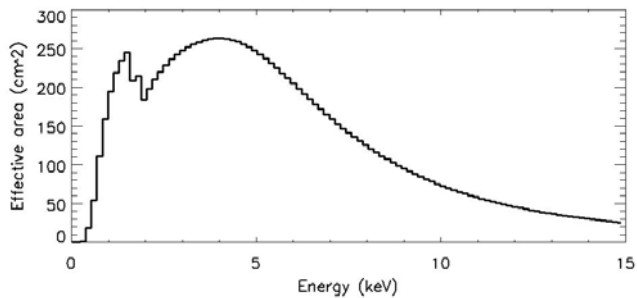

**Figure 25** Calculated total on-axis effective area of LE (including both the narrow and wide FOVs SCDs).

### 4.5 Time response

The distribution of time intervals between two adjacent events for a single detector channel is shown in Figure 26. The vertical green line represents the intervals of forced triggers (see section 3), which is 32 ms for each channel. The blue line represents the intervals of the single events, which follows exactly an exponential distribution, while the red line shows the intervals of all events.

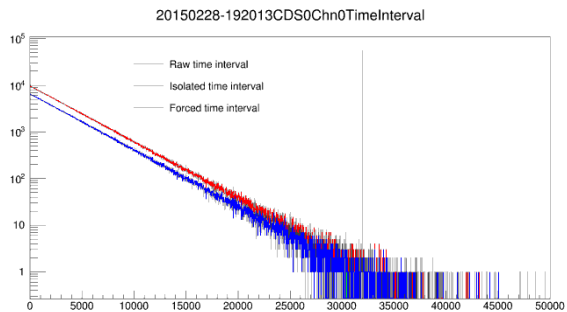

**Figure 26** Statistical distribution of time interval between two events for a single detector channel.

Since the charge packets in the pixels are transferred one by one, the time responses of LE depend strongly on the charge transfer process, which can be obtained well under the long exposure mode (see section 3). In this mode, the driving clock stops during the exposure time (typically 10 ms). During the next 150 driving clocks (10 μs per driving clock), the signals are read out continuously.

One can obtain the time response by folding the readout signal time with a period of 11.5 ms (10 ms of exposure plus 1500 μs of transfer) as showed in Figure 27. In principle, the response is proportional to the area of each L-shaped electrode. There is a flat part at the beginning of the response, which is due to the 20 small pixels close to the readout anode in the CCD236. There is also a flat part in the end with 32 driving clocks (320 μ s). Therefore, the number of effective pixels for time response is 98, i.e. 150-20-32 = 98. For the LE sensor, the area of the electrode increases with its distance to the readout end, and so there are more signal counts as the readout time increases. The last 32 signals are caused by the X-ray photons arrived during the transfer period. Therefore, the practical time resolution is 0.98 ms. These two dips in Figure 27 are caused by the forced triggers of other detectors, in which time the signals cannot be recorded according to the design of the electronics.

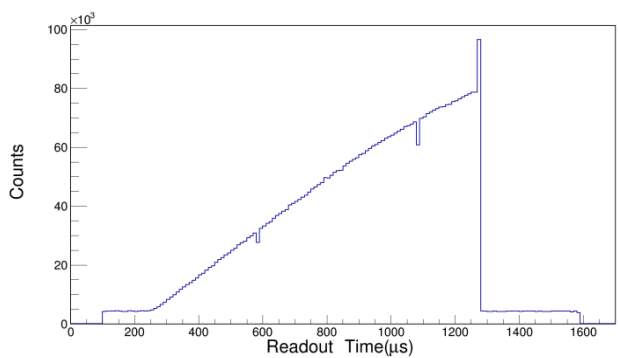

**Figure 27** Time response of LE obtained under long exposure mode.

Although the recorded readout time follows a triangle-like distribution strictly as shown in Figure 27, the exact TOA of each photon cannot be obtained more accurately than 0.98 ms.

### 4.6 Other aspects

#### 4.6.1 Absolute Time

LE acquires the absolute time (AT) by recording both GPS second signal and its corresponding UTC time (the counting of the second relative to a particular starting time) from the platform. The accuracy of the GPS second signal is better than 10 μs, and the corresponding UTC time is a broadcasting message from the LEB through the 1553B bus.

The AT acquisition is conducted as follows, after receiving the GPS second pulse and UTC instructions, a relative time information is added to form the GPS event and UTC event which are both sent to LEB through the LVDS interface. The GPS event and UTC event are added into the event data (the former is added to the science data packets, the latter to the engineering data packet) which is sent to the platform by LEB.

Under normal circumstances, each LED operates under a high-precision 5 MHz clock from the platform; however, it automatically switches to an internal clock for timing if this high-precision clock fails.

### 4.6.2 Pileup

Since the charge transfer in CCD236 and the readout of charge are continuous, the fraction of pileup events is less than 1% even for the brightest sources such as Sco X-1, whose maximum count rate is about 18000 cts/s, such a pileup fraction is much lower than that of focusing telescopes, e.g. XMM-Newton [22]. The comparison of the pileup among LE and XMM-Newton MOS and PN in small window mode is shown in Figure 28.

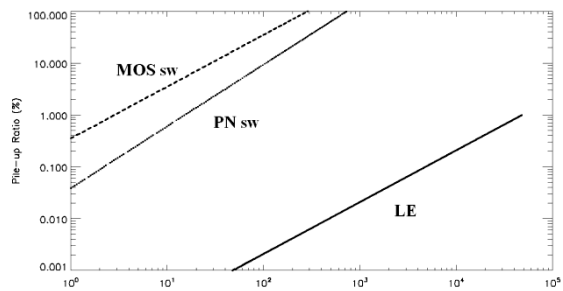

**Figure 28** The pileup of LE compared to the XMM-Newton MOS and PN in small window mode.

## 5 Conclusions

In this paper, we have introduced the main components of LE onboard Insight-HXMT, including the mechanical structures, electronic diagram and functions, and the performance specifications of the instruments. It is shown that LE has a total geometrical area of 384 cm2, contains different FOVs that can be used to determine the diffuse X-ray and particle induced background components. LE achieves a good energy resolution (160 eV@8 keV), fairly good time resolution (0.98 ms) and extremely low pileup (<1%@18000 cts/s), which makes it a unique and powerful telescope to observe bright X-ray sources, e.g. neutron star and black hole X-ray binaries in outbursts.

The Insight-HXMT has launched in 2017 and LE works well since then. The on-orbit performance of LE meets the requirements of the design [2, 21, 23-25] and several science results [26-29] have already been obtained.

*The author expresses their thanks to the people helping with this work, and acknowledges the valuable suggestions from the peer reviewers. This work was supported by the Strategic Priority Research Program on Space Science, the Chinese Academy of Sciences, Grant No.XDA040102.*

## 7 References

1    Y. Chen, W.W. Cui, Modern Physics, 25, 4 (2016) (in Chinese).

2    Y. Chen, W.W. Cui, W. Li, et al. Spacecraft Engineering, 27, 134 (2018) (in Chinese).

3    B. Aschenbach, U. Briel, F. Haberl, et al. SPIE, 4012, 731 (2000).

4    X.Y. Liu, B. Lu, Y.J. Yang, et al. CPC (HEP&NP), 39, 076101 (2015).

5    B. Lu, W.W. Cui, Y.S. Wang, et al. CPC (HEP&NP), 36, 846 (2012).

6   Y.P. Chen, S. Zhang, J.L. Qu, et al. ApJ, 864, 30 (2018).

7   Y. Huang, J.L. Qu, S.N. Zhang, et al. ApJ, 866, 122 (2018).

8   T.P. Li, M. W, Z. G. Lu, et al. Ap&SS, 205, 381 (1993).

9   F. J. Lu, T. P. Li, X. J. Sun, et al. A&AS, 115, 395 (1996).

10  Y.S. Wang, Y. Chen, Y.P. Xu, et al. CPC (HEP&NP), 34, 1812 (2010).

11  B.G. Lowe, A.D. Holland, I.B. Hutchinson, et al. NIM A 458, 568 (2001).

12  A.D. Holland & P. Pool, Proc SPIE 7021, 702117 (2008).

13  Y.S. Wang, Y. Chen, Y.P. Xu, et al. CPC (HEP&NP), 36, 991 (2012).

14  Y.S. Wang, Y.J. Yang, Y. Chen, et al. CPC (HEP&NP), 38, 066001 (2014).

15  Rothschild, R., Boldt, E., Holt, S. et al. Space Science Instrumentation, 4, 269 (1979)

16  W. Hu, T.X. Chen, Q.R. Sun, et al. Mesh. Applied Mechanics and Materials, 290, 21 (2013).

17  W. Hu, T.X. Chen, Y. Chen, et al. Fabrication of a multilayered optical device based on submicron polyimide film. In:
    International Conference on Materials Science, Energy Technology, Power Engineering. Atlantis Press, 2016

18  X. Cao, W. Jiang, B. Meng, et al. 2019arXiv191004434 (2019)

19  S. Zhang, Y.P. Chen, Y.N. Xie, et al. SPIE.9144E.55Z (2014)

20  X.Y. Liu, Y.J. Yang, Y. Zhu, et al. Nuclear Electronics & Detection Technology, 36, 144 (2016) (in Chinese).

21  X.B. Li, M.Y. Ge, X.F. Li, et al. Spacecraft Engineering, 27, 143 (2018) (in Chinese).

22  P. Jethwa, http://xmm2.esac.esa.int/docs/documents/CAL-TN-0200 -1-0.pdf (2012)

23  X.B. Li, L.M. Song, X.F. Li, et al. SPIE, 10699, 69 (2018)

24  X.B. Li, L.M. Song, X.F. Li, et al. 2019arXiv191004390 (2019)

25  X.F. Zhao, Y.X. Zhu, D.W. Han, et al. 2019arXiv191004702 (2019)

26  Y.P. Chen, S. Zhang, J.L. Qu, et al. ApJ 864, 30 (2018)

27  Y. Huang, J.L. Qu, S.N. Zhang, et al. ApJ, 866,122 (2018)

28  Y. Zhang, M.Y. Ge, L.M. Song, et al. ApJ, 879, 61 (2019)

29  S.J. Zheng, S.N. Zhang, F.J. Lu, et al. ApJS, 244, 1(2019)

Abbreviation

| | |
|---|---|
| AMS | All Sky Monitoring |
| AT | Absolute Time |
| CCD | Charge Coupled Device |
| DMM | Data Management Module |
| EBIT | Electron Beam Ion Trap |
| FOV | Field of View |
| FT | Forced Trigger |
| FWHM | Full Width Half Magnitude |
| HPGe | High Purity Germanium |
| HXMT | Hard X-ray Modulation Telescope |
| INL | Integral Linearity |
| LE | Low Energy X-ray Telescope |
| LEB | LE Electric Control Box |
| LED | LE Detector Box |
| LVDS | Low Voltage Differential Signaling |
| OBF | Optical Blocking Filter |

| OTT | Over Threshold Trigger |
|-----|------------------------|